\documentclass[conference]{IEEEtran}
\usepackage{cite}
\usepackage{amsmath,amssymb,amsfonts}
\usepackage{algorithmic}
\usepackage{graphicx}
\usepackage{textcomp}
\usepackage{xcolor}
\def\BibTeX{{\rm B\kern-.05em{\sc i\kern-.025em b}\kern-.08em
    T\kern-.1667em\lower.7ex\hbox{E}\kern-.125emX}}

\usepackage{xcolor}
\usepackage{amsfonts} % For mathbb.
\usepackage{amsmath} % for \mod and matrices.
\usepackage{nicefrac} % For nicefrac.
\usepackage{amssymb} % For triple-barred implications.
\usepackage{comment} % For comment environment.
\usepackage[shortlabels]{enumitem}% For letter enumeration
\usepackage{wasysym} % For \smiley and \hexagon.

\usepackage{nicematrix}

\newcommand\Frownie {\frownie}
\newcommand\QIF {{\sf QIF}}
\newcommand\HMM {{\sf HMM}}
\newcommand\Kuifje {{\sf Kuifje}}
\newcommand\Kuifjetje {{\small\sf Kuifje}}
\newsavebox{\KuifjeBOX}\sbox{\KuifjeBOX}{\Kuifje}
\newsavebox{\KuifjeBOx}\sbox{\KuifjeBOx}{{\small\Kuifje}}
\newcommand\Section[1] {\section{#1}}
\newcommand\Subsection[1] {\subsection{#1}}
\newcommand\Subsubsection[1] {\subsubsection{#1}}
\newcommand\Label[1] {\label{#1}{~\mbox{\tiny[#1]}}}
\renewcommand\Label[1] {\label{#1}}
\newcommand\Assign[2]{\mbox{$#1$\,:=\ $#2$}}
\newcommand\Exp[1][] {\ensuremath{\ifx\empty#1\empty\mathit{exp}\else\mathit{#1}\fi}}

\newcommand\Eqv {\ensuremath{\equiv}}
\newcommand\If {\textsf{if}}
\newcommand\Then {\textsf{then}}
\newcommand\Else {\textsf{else}}
\newcommand\While {\textsf{while}}
\newcommand\Skip {\textsf{skip}}
\newcommand\Do {\textsf{do}}
\newcommand\Od {\textsf{od}}
\newcommand\Print {{\sf print}}
\newcommand\Printje {{\small\sf print}}
\newcommand\Fig[1] {Fig.\,\ref{#1}}
\newcommand\FIG[1] {Figure\,\ref{#1}}
\newcommand\Sec[1] {Sec.\,\ref{#1}}

\newcommand\App[1] {App.\,\ref{#1}}

\newcommand\Eqn[1] {(\ref{#1})}
\newcommand\EqnSmall[1] {\textrm{\small(\ref{#1})}}
\newcommand\EQn[1] {Eqn.\,(\ref{#1})}

\newcommand\Prog[1] {Prog.\,\ref{#1}}

\newcommand\Dist {\ensuremath{\mathbb D}}
\newcommand\Pow {\ensuremath{\mathbb P}}
\newcommand\Hyp {\ensuremath{\mathbb D^2}}

\newcommand\Supp[1] {\ensuremath{\lceil#1\rceil}}
\newcommand\Fun {\ensuremath{\mathbin\rightarrow}}
\newcommand\EV[2] {\ensuremath{{\cal E}_{#2}{#1}}}
\newcommand\EVE[3] {\ensuremath{({\cal E}#1{:}\,#2 \cdot #3)}}
\newcommand\PLUS {\ensuremath{\mathbin{\fbox{\tiny$+$}}}}
\newcommand\MAX {\ensuremath{\mathbin{\fbox{\tiny$\sqcup$}}}}
\newsavebox\MAXBox\sbox{\MAXBox}{\MAX}
\newcommand\MAx {\ensuremath{\bigsqcup}}
\newcommand\AND {\ensuremath{\mathbin{\fbox{\tiny$\times$}}}}
\newcommand\Max {\ensuremath{\mathbin{\sqcup}}}
\newcommand\NF[2] {\ensuremath{\nicefrac{#1}{#2}}}
\newcommand\IB[1] {\ensuremath{[#1]}}
\newcommand\False {\textsf{false}}
\newcommand\True {\textsf{true}}
\newcommand\Q[1] {\makebox[0pt][l]{\quad\mbox{\rm#1}}}
\newcommand\Wide[1] {\quad#1\quad}
\newcommand\WIDE[1] {\Wide{\Wide{#1}}}
\newcommand\GFQ[3] {\ensuremath{(\,\MAX\,#1\mid#2\cdot#3\,)}}
\newcommand\GfQ[2] {\ensuremath{(\,\MAX\,#1\cdot#2\,)}}
\newcommand\QQ[4] {\ensuremath{(#1\,#2\mid#3\cdot#4)}}

\newcommand\Subst[3] {\ensuremath{#1[#2\backslash#3]}}
\newcommand\Mod {\ensuremath{\mathbin{\textsf{mod}}}}
\newcommand\Div {\ensuremath{\mathbin{\textsf{div}}}}
\newcommand\Not[1] {\ensuremath{\overline{#1}}}
\newcommand\Cmt[1] {//~#1}
\newcommand\Cents {\mbox{\textcent}}
\newcommand\Gpg {\textsf{gpg}}
\newcommand\GPg[1] {\ensuremath{\Gpg.#1}}
\newcommand\GPG[2] {\ensuremath{\Gpg.#1.#2}} % Alternative is \Gpg(#1,#2). Both are used.
\newcommand\In {{:}\,} % or {{\in}}

\newcommand\CNote[1] {\textcolor{blue}{#1}}
\newcommand\Comp {\ensuremath{\mathbin\circ}}
\newcommand\MDP {{\small\sf MDP}}

\newenvironment{Reason}{\begin{tabbing}\hspace{2em}\= \hspace{1cm} \= \kill}
    {\end{tabbing}\vspace{-1em}}
\newcommand\Step[2] {#1 \> $\begin{array}[t]{@{}llll}#2\end{array}$ \\}
\newcommand\StepR[3] {#1 \> $\begin{array}[t]{@{}llll}#3\end{array}$
    \` {\RF \makebox[0pt][r]{\begin{tabular}[t]{r}``#2''\end{tabular}}} \\}

\newcommand\Space {~ \\}
\newcommand\RF {\small}

\newcommand\Avg {\ensuremath{\mathop\mu}}
\newcommand\Unit {\ensuremath{\mathop\eta}}

\newcommand\HRef {\ensuremath{\mathbin\sqsubseteq}}
\newcommand\Real {\ensuremath{\mathbb R}}
\newcommand\Zero {\ensuremath{\mathbin{\textsf{0}}}}

\newtheorem{Definition}{Definition}

\newtheorem{Lemma}[Definition]{Lemma}

\newcommand\LEM[1] {Lemma \ref{#1}}

\begin{document}
\title{Source-level reasoning \\ for quantitative information flow}

\author{
\IEEEauthorblockN{Chris Chen}
\IEEEauthorblockA{\textit{School of Computing} \\
\textit{Macquarie University}\\
Australia \\
chris.chen1@mq.edu.au}
\and
\IEEEauthorblockN{Annabelle McIver}
\IEEEauthorblockA{\textit{School of Computing} \\
\textit{Macquarie University}\\
Australia \\
annabelle.mciver@mq.edu.au}
\and
\IEEEauthorblockN{ Carroll Morgan}
\IEEEauthorblockA{\textit{School of Engineering and Trustworthy Systems} \\
\textit{UNSW}\\
Australia \\
carroll.morgan@cse.unsw.edu.au}}

\maketitle
\begin{abstract}
We present a novel formal system for proving quantitative-leakage properties of programs. Based on a theory of Quantitative Information Flow (QIF) that models information leakage as a noisy communication channel, it uses ``gain-functions'' for the description and measurement of expected leaks.

We use a small imperative programming language, augmented with leakage features, and with it express adversaries' activities in the style of, but more generally than, the Hoare triples or expectation transformers that traditionally express deterministic or probabilistic correctness but without information flow.

%We annotate the programs
The programs are annotated
with ``gain-expressions'' that capture simple adversarial settings such as ``Guess the secret in one try.''\ but also much more general ones; and our formal syntax{\boldmath$+$}logic -based framework enables us to transform such gain-expressions that apply \underline{after} a program has finished to ones that equivalently apply \underline{before} the program has begun.

In that way we enable a formal proof-based reasoning system for QIF
at the source level.
We apply it to the  %programming 
language we have chosen, and demonstrate its effectiveness in a number of small but sometimes intricate situations.
\end{abstract}


\Section{Introduction\Label{s1637}}
In 2009 Geoffrey Smith proposed a radical new approach for evaluating the impact of information leaks in security-sensitive programs \cite{Smith:2009aa}. {\bf Our contribution here} is to present an approach for using his ideas at the \emph{source-code level} of to-be-secured systems, rather than (only) at the level of mathematical models of them: we are aiming at a \emph{p\underline{ro}g\underline{ram} \underline{lo}g\underline{ic} of quantitative information flow}.

We begin with a review of the role of ``quantitative''. 

Well before Smith's 2009 work, the prevailing approach based on ``absolute'' non-interference security \cite{Goguen:84} had begun to be been seen as too stringent to allow many practical, indeed necessary, systems to be considered secure. Thus quantifying the ``amount'' of information leaked --introducing the concept of  \emph{Quantitative Information Flow}, or \QIF-- had already begun to be investigated (e.g.\ by Clark, Hunt  and Malacaria et al.\ \cite{ClarkHM01,ClarkHM05,DBLP:journals/jcs/ClarkHM07}  and others \cite{Sabelfeld:01,Kopf:07}); and it was the ``{\sf Q}'' in \QIF\ that enabled the more nuanced criterion of ``secure \emph{enough}''. Geoffrey Smith's award-winning\,%
%\footnote{\ToDo{URL to ``Test of Time'' award}}\
paper [op. cit.]\ --building on that earlier work-- finally made \QIF\ an essential advance rather than merely an interesting possibility.

An information-flow analysis views a program as a system that processes ``high security'' secret information, e.g.\  passwords; but sometimes programs' treatment of secrets can be correlated to observed behaviour, or other data of ``low security''. Those programs then ``leak'' information when an adversary can infer something about the ``high security'' secret by using their knowledge of the program code, and observing real-time behaviours such as control flow, timings, or low-security correlations. But ``with the {\sf Q}'', instead of \emph{banning} such correlations altogether, we can \emph{measure} them: and those measurements then become accessible to formal analysis. But what are the best measurement techniques to use?

Early models proposed traditional Shannon-style information theory --- there, the secret is modelled as a probability distribution over its potential values, and a joint distribution between the secret and real-time observables can then be created and analysed using, for example, conditional Shannon Entropy. But, going against the flow, in a series of compelling examples Smith showed that those Shannon-style measurements were unable to explain whether a leak was actually damaging. Here is one of Smith's examples \cite[pp.294ff]{Smith:2009aa}.

%Using $H$ for a secret (hidden) variable of $k$ bits, and $L$ for an observable (visible)
Using $H$ for a secret (hidden) variable of $8k$ bits, and $L$ for an observable (visible) variable, whenever $H$ is divisible by 8 the program
\begin{align}
	\text{\If\ $\;H\Mod 8 = 0\;$ \Then\ \Assign{L}{H}\ \Else\ \Assign{L}{1}} \Label{e1405}
\end{align}
assigns the entire value of hidden $H$ to visible $L$ (i.e.\ delivers it wholly to the adversary). But the program
\begin{align}
	\Assign{L}{\quad H\;\;\&\;\; 0^{7k{-}1}1^{k{+}1}}\hspace{10em}\Label{e1411}
\end{align}
leaks only the $k{+}1$ low-order bits of $H$ 
(where ``$\&$'' is bitwise {\sf\small and}). Smith calcuated that, for various reasonable settings of $k$, conditional Shannon entropy (${\sf H}(H{\mid}L)$) measurements of those programs rate \Eqn{e1411} \underline{sli}g\underline{htl}y \emph{less} secure than \Eqn{e1405}.
That notwithstanding, he argued that \Eqn{e1411} should be considered \underline{substantiall}y \emph{more} secure than \Eqn{e1405} --- an adversary's knowing ``all of the secret some of the time'' is far more damaging than her knowing ``some of the secret all of the time''.
%What Smith argued was missing 
Missing from the Shannon entropy approach was the \emph{intent} of the adversary; and reflecting that intent is what focusses the analysis on damaging scenarios.  Smith proposed Bayes' vulnerability as an important such measurement because it shows that \Eqn{e1405} is indeed much more problematic than \Eqn{e1411} in a security scenario where the intent of the adversary is to  guess the secret.

Smith's novel proposal eventually led to the now-influential \emph{g-leakage framework} \cite{Alvim:2012aa}, where a program is modelled mathematically as a noisy information channel (i.e. a stochastic matrix taking secrets to leaks); and adversarial contexts are modelled complementarily as sets of ``actions'', each one a random variable over the (posterior) distributions induced by the adversary's observations of those leaks. Each random variable gives an expected gain (for the adversary) should she take that action on the deduced posterior.

A typical ``run'' (modelled in the above mathematical way) begins with a prior distribution (known to the adversary) on the input, and a communication channel (as above, also known) that, for each possible output, determines a corresponding ``revised'' posterior distribution on the input. The (rational) adversary then chooses, from her set of actions, one whose expected value over the posterior input distribution (which she calculates from the prior, the channel and the output she saw) is the greatest. That is her reward; and she takes that action.

Of course the adversary does not know beforehand which output she will observe, but from the prior and the program text she does know the \emph{distribution} of those outputs. So in one further mathematical step she calculates the expected ``greatest value'' (as above) over the distribution of possible outputs: that is the overall vulnerability represented by that prior, that channel and that adversary. That number is the \emph{vulnerability} of that system overall: that prior, that channel and that adversary.

The ideas above have led to a number of analysis tools \cite{LibQif,FoPPS:19}, and reasoning based on them has been used to investigate many \QIF\ examples in security and privacy \cite{Meng:2011aa,KopfS10,JuradoP021} --- but those tool-based approaches usually rely on mathematical models of the specific (and  \emph{finite}) scenarios to be investigated. What they typically do \emph{not} do is to allow \QIF\ analysis \emph{at the source-code level} of the program --- that is analyses based on logical proofs in which a program's actual source code is used directly to show that an attack's expected damage is limited,
%will not succeed, 
or to find the vulnerable points in a design which an attacker can use to her advantage.

{\bf Thus in this paper we study a \emph{generalisation} of traditional source-level program logics} --i.e.\ assertion-based formal reasoning-- that can describe and \emph{quantifiy} both program leaks and the adversarial contexts in which they might be used. In particular the focus is on describing adversarial ``successes'' quantitatively, and showing how to estimate them by examining the initial state (i.e.\ prior) distribution, and the program even before it is run. 

{\bf A summary of our attack model is given in \Sec{s1627}}.

We augment \Fig{e1405} above with program-logic style ``assertions'' that express what we have --for now-- added as comments in \Fig{e1228} below: an adversary guesses the value of ``high security'' $H$ by examining ``low security'' $L$ after executing \Eqn{e1405}.

\begin{equation}\Label{e1228}
\vspace{1em}
\begin{tabular}{l}
	\Cmt{Attack will succeed if $H\Mod 8 = 0$. ($\dagger$)} \\
%\# otherwise only if can guess any other value for H  \\[1ex]
	\text{\If\ $\;H\Mod 8 = 0\;$ \Then\ \Assign{L}{H}\ \Else\ \Assign{L}{1}} \\
	\Cmt{Attack: Guess the  value of $H$.} \\
\end{tabular}
\end{equation}

%A similar annotation for \Prog{e1411} would indicate that only $7k{-}1$ bits of the input bits remain protected, and which of those bits they are.  

The particular strength, though,  of understanding vulnerabilities in this source-based way is that it facilitates compositional reasoning (in the same way that standard assertion-based reasoning does). For example, when a program fragment operates within a larger context we can use the fragment's annotations as a \emph{specification}, to understand the vulnerabilities of that larger context that are caused by the smaller fragment within it. For example if \Eqn{e1405} were to execute  \emph{after} the assignment  \Assign{H}{H{*}8}, we would know immediately from the annotated version \Eqn{e1228} of \Eqn{e1405} that the whole program is vulnerable, because \Assign{H}{H{*}8} establishes exactly the conditions under which the final attack will succeed --- we can read them off from ($\dagger$) in \Eqn{e1228}.

\textbf{Our specific contributions are these:}

\begin{enumerate}
\item Based on the $g$-leakage approach, we show how to describe adversarial attacks as a generalisation of assertions expressed as predicates (or expectations \cite{McIver:05a}) on the program variables, formalising the descriptions just above: we call such expressions \textbf{gain expressions}. They are the basis for \textbf{leakage specifications}, giving vulnerabilities on inputs corresponding to attacks on outputs.
\item\Label{i1405} We describe a small source-level reasoning system which enables a posterior gain expression describing the adversary's intent  to be \textbf{transformed} to an equivalent pre-gain expression describing exactly the prior knowledge of the adversary that motivate their posterior attack.
\item We illustrate our approach on a number of examples.
\end{enumerate}

\textbf{Overall our results} represent a step towards a proof-based rather than experiment-based analysis of Quantitative Information Flow for the important application of concrete implementations of security systems.

\Section{The \usebox{\KuifjeBOX} language\Label{s1340}}
\Subsection{Syntax\Label{s1426}}
\begin{figure}
\begin{tabular}{r@{\qquad}p{15em}}
\textit{skip}
	& \Skip
	\\& Do nothing. \\[1ex]
\textit{assignment}
	& \Assign{x}{\Exp}
	\\& Assign expression \Exp\ to variable \Exp[x]. \\[1ex]
\textit{composition}
	& \Exp[stmt1] ;\,$\cdots$\,; \Exp[stmtN]
	\\& Execute \Exp[stmt1] to \Exp[stmtN] in order. \\[1ex]
\textit{conditional}
	& \If\ \Exp[cond\,]~~\Then\ \Exp[stmtT]~~\Else\ \Exp[stmtE]
	\\& Evaluate Boolean \Exp[cond];\newline
	    execute \Exp[stmtT] if true,\newline
	    excute \Exp[stmtE] otherwise.\newline
	    A missing \Else\ is ``\Else\ \Skip''. \\[1ex]
\textit{loop}
	& \While\ \Exp[guard\,]\ \Do\ \Exp[body] \Od
	\\& Evaluate Boolean \Exp[cond];\newline
	    execute \Exp[body] if true, then repeat.\newline
	    Exit otherwise. \\[1ex]
\textit{leak}
	& \Print\ \Exp
	\\& Evaluate \Exp,\newline
	    and leak its value to the adversary.
\end{tabular}

\bigskip
The syntactic novelty of \Kuifje\ is the \Print\ statement. All the variables in the state space are hidden from the adversary, that is they are all ``high level'' (\Sec{s1249}); but she is made aware explicitly of expressions ``leaked'' (via \Print). Those expressions are program-level descriptions of leaks we already know the program to have. The {\sf Q} in \QIF\ is what allows us to determine whether they are tolerable.

A semantic novelty (\Sec{s1357}) is that information is \emph{always} leaked ``implicitly'' by conditional branches, whether \If's or \While's.
\caption{Basic syntax of \Kuifje\Label{f1341}}
\end{figure}

In \Fig{f1341} is (a portion) of the syntax of \Kuifje, a small \QIF-aware programming language \cite{FoPPS:19,Alvim:20a},%\ToDo{also %\Cite{JuradoP021}? \Cite{anyone else}?} 
which includes the usual basic imperative constructions \emph{plus} a novel \Print\ statement which leaks information explicitly to the adversary.

We do not include variable declarations or local scopes in this presentation: all variables here will be global \emph{and hidden}, with their types evident from context.

\Subsection{Semantics of \usebox{\KuifjeBOx} (informally)\Label{s1427}}

\Subsubsection{Review of \usebox{\KuifjeBOx}'s antecedents\Label{s1451}}~\\
A conventional deterministic sequential program operates over a state-space $S$ say, taking an intial state $s$ (in $S$) to some final state $s'$ (also in $S$). (We consider only terminating programs here, and so do not need a $\bot$ or similar for non-termination.)

Sequential programs with demonic choice \cite{Dijkstra:76} take initial $s$ to some (possibly singleton) subset set $S'$ of $S$, and the choice of $s'$ in $S'$ that the execution reaches is unpredictable.

Programs with probabilistic- instead of demonic choice \cite{Kozen:83} take initial $s$ to some (possibly point) distribution $\delta$ on $S$, and the choice of final state $s'$ from the support \Supp{\delta} of that $\delta$ is made according to a distribution that the program determines. (With finite state-spaces, our distributions will be discrete.)

Programs with \emph{both} probabilistic and demonic choice \cite{McIver:05a} take initial $s$ to some (possibly singleton) set of (possibly point) distributions $S$. First the choice of distribution $\delta$ is made demonically; then the choice of final $s'$ from that \Supp{\delta} is made probabilistically.

\Subsubsection{\usebox{\KuifjeBOx}'s ``extra step'' for information flow\Label{s1424}}
\Kuifjetje\ programs take an initial \emph{distribution} $\pi$ on the state-space $S$ (rather than an initial state) to a final distribution $\Delta$ \emph{of distributions} on $S$ (rather than of states in $S$), thus of type $\Dist\Dist S$ --- equivalently written $\Hyp S$. Members of $\Hyp S$ are called \emph{hyper-distributions} or, for short, just ``hypers''.

The intuition supporting the above model is that the adversary knows the prior distribution $\pi$ on $S$, but does not know the actual value of $s$ on any particular run. As the program runs, the adversary follows the program counter and --knowing the source code-- she \emph{revises} that prior $\pi$ to some (other) posterior $\delta$, using Bayesian reasoning, based on any leaks she observes; and if there are several leaks, there will be several revisions.

At the end of the the program's execution, she will have a final revised posterior; but she cannot predict \emph{before} the program runs which observations she will see and --therefore-- she also cannot predict which posterior she will finally have. But she \emph{can} predict the distribution of those possible posteriors --- and that distribution of posteriors (themselves distributions) is the hyper $\Delta$ that the semantics gives for the effect of running the (known) program on the (known) prior $\pi$.

Thus the ``operational'', forward semantics of a \Kuifjetje\ program has type $\Dist S \Fun \Hyp S$.

\Section{Source-level reasoning in \usebox{\KuifjeBOX}\Label{s1431}}

\Subsection{What is ``source-level reasoning'' in this context?\Label{s1718}}

All of \Kuifjetje's antecedents mentioned in \Sec{s1451} above have operational semantics (they were in order $S\!\Fun\!S$, $S\!\Fun\!\Pow S$, $S\!\Fun\!\Dist S$ and $S\Fun\Pow\Dist S$); but even the original $S\!\Fun\!S$ can be tedious to use on actual source code: the program text must be ``converted'' to a relation first. A major step therefore in reasoning about actual specific programs was Hoare-logic \cite{Hoare:69}, based on Floyd's introduction of assertion-annotated flowcharts \cite{Floyd:67}. And Hoare's logic was later extended to see a program as predicate-transformer ``going backwards'' --- that is from a postcondition that the program was to achieve to the weakest precondition that would ensure it would do so \cite{Dijkstra:76}. That backwards approach was later extended to the probabilistic models as well \cite{Kozen:83,McIver:05a}, replacing pre- and post-\underline{conditions} with pre- and post-\underline{ex}p\underline{ectations}.

Here we will extend the backwards approach further, to \Kuifje, generalising ``expectations'' to ``gain expressions''.

\Subsection{Expectations generalise assertions\Label{s1715}}
Mathematically, \emph{Expectations} are non-negative real-valued functions $f$ say on the program state \cite{Kozen:83,McIver:05a}, generalising assertions by taking 0 to \False\ and 1 to \True. (The generalisation is that he expected value of a characteristic function of an assertion is the probability of the assertion's truth.) If the final distribution of the program's state is some $\delta$, then $f$ might express that one hopes to gain the expected value of $f$ over that distribution $\delta$ once the program has run, just as one hopes that a program's postcondition will be true,\,%
\footnote{Taking $f$ and $\delta$ together we have a ``random variable'' over $S$.}
and we write \EV{f}{\delta} for that. Alternatively $f$ can be written as a ``gain expression'' \Exp\ containing free program variable(s) $x$ say, i.e.\ at the source level so that $f(x){=}\Exp$, and then we could equivalently write \EVE{x}{\delta}{\Exp} for \EV{f}{\delta}, with the free $x$'s in \Exp\ being bound by the quantifier \EVE{x}{\delta}{-}. 

\Subsection{Gain functions/expressions generalise expectations\Label{s1722}}~\\
\vspace{-4ex}
\Subsubsection{What is a gain function?\Label{s1520}}
A \emph{gain function} is a \emph{set} of expectations, and it is having a set of them (rather than just one) that enables us to model the possible \emph{actions} of a rational adversary. Each expectation in the gain function corresponds to one of her possible actions; rationality means ``choosing the best action for her, given what she knows about the prior and the program, and the observations she has made.

Suppose for example the adversary is trying to guess the value of a state-variable $0{\leq}x{\leq}9$ which she knows to be uniformly distributed over its ten possible values: her ten possible actions are then ``guess that $x$ is 1'', ``guess 2'' etc.\ and let's say for now that she gains \$1 if her guess is correct, and \$0 otherwise.\,%
\footnote{There is a great variety of gains possible in general, much more expressive than simply ``Guess and get \$1 if you are right \cite{Alvim:20a,Alvim:2012aa}.}

Which action maximises her expected gain?

Without further information (than that the prior is uiform), each of her possible actions yields the same expected gain, that is 10\Cents. (It is equivalently seen as the probability \NF{1}{10} that she has guessed correctly, because a correct guess yields \$1: recall that the probability of an event is the expected value of its characteristic function.)
%\footnote{\ToDo{Emphasise this?}}\,)
But now suppose further that a program, operating on that prior, leaks whether $x$ is a multiple of 3, and the adversary is able to observe that leak. If she observes ``$x$ is a multiple of 3'' then \emph{rationally} she will guess one of $0,3,6,9$ and her expected gain in that case will be \$\NF{1}{4}; if she observes ``isn't,'' she will instead guess one of $1,2,4,5,7,8$ and her expected gain will be \$\NF{1}{6}.

The essential detail above is that the adversary \emph{makes a rational choice} based on what she knows, and what she has seen: ``Which number should I guess to maximise my expected gain?'' But she cannot make that choice until the program is run, and her approach to determining her \emph{overall} expected gain is then based on the following informal (but correct) reasoning:
\begin{quote}
Because the prior is uniform, I know (without even running the program) that I will observe ``is'' with probability \NF{4}{10}, and ``isn't'' with probability \NF{6}{10}.
\begin{itemize}
\item Thus in \NF{4}{10} of the runs I will guess 3 (or 0 or 6 or 9), and expect to gain 25\Cents; but
\item in \NF{6}{10} of the runs I will guess (say) 8, and expect to gain \$\NF{1}{6}.
\end{itemize}
My expected gain is thus \(\NF{4}{10}{\times}\NF{1}{4}+\NF{6}{10}{\times}\NF{1}{6} = 20\Cents\), i.e.\ overall
twice as much as the 10\Cents\ it would have been had I not run the program.\,%
\footnote{The ``twice'' is because the \Print\ statement divided the state-space into two pieces: see \cite[Thm.\,1.1]{Alvim:20a}.}
\end{quote}
That her expected gain \emph{after} running the program \emph{is more than it was before she ran it} --in fact twice as much-- is how \QIF\ tells us that the \Print\ has made the program insecure.

Our aim with \emph{source-level reasoning} to be able to reach that same conclusion but by working with program source-code and gain functions written as gain \emph{expressions} over the program variables, as suggested in \Fig{f1645}, rather than via informal reasoning as we did just above: that kind of reasoning is too unreliable for programs of any size or complexity.
\begin{figure}
\begin{tabular}{l}
\# Initial gain expression,
   evaluated over the prior on $x$. \\[1ex]
\Print~~$(x\Mod3){=}0$ \\[1ex]
\# Final gain expression,
   expressing adversary's actions \\
\# in terms of guesses of $x$.
\end{tabular}

\bigskip
How do we convert the final gain expression into the initial gain expression
using source-level program logic?
\caption{Guessing a number between 1 and 10\Label{f1645}}
\end{figure}
\Subsubsection{What syntax do we use for gain expressions?\Label{s1716}}
Gain expressions are written either as an explicitly enumerated set of expectations, if that set is small (or even singleton), or as a quantification over a new operator \MAX\ introduced for that purpose. The expectations themselves are (syntactically) non-negative real-valued expressions over the program variables (\Sec{s1715}). Those expectations are often based on Boolean (rather than intrinsically real-valued) expressions, thus we use the convenient convention that \IB{\False} is 0 and \IB{\True} is 1.\,%
%\footnote{These ``Iverson brackets'' were used extensively in \cite{McIver:05a} for just that purpose.}

Each component (i.e.\ real-valued expression) of a gain expression gives the random variable (a function over the program's state) whose expected value accrues to the adversary should she decide to take the action corresponding to that component. In that way, the expectations abstract from the actual actions: we are not interested in what those actions are; we care only what (expected) benefit each one might bring to our adversary. Being rational, she will choose from her gain expression --her set of expectations-- the action whose expected value is (perhaps equally) greatest over the distribution she believes the state variables to have (a posteriori); and that belief is formed by (Bayesian) revision of the program's initial prior distribution and the observations she makes as the program runs.

In the example of \Sec{s1520}, the potential actions of the adversary were ``guess that $x$ is 0'', ``guess that $x$ is 2'',\ldots, ``guess that $x$ is 9''; and she would receive \$1 if her guess was correct (and \$0 otherwise). The expectations for each of those actions are resp.\ \IB{x{=}0}, \IB{x{=}1},\ldots, \IB{x{=}9}, and so the gain expression as a whole for this adversary is enumerated
\begin{equation}\Label{e1024}
	\IB{x{=}0} \,\MAX\, \IB{x{=}1} \,\MAX\, \cdots \,\MAX\, \IB{x{=}8} \,\MAX\, \IB{x{=}9} \end{equation}
having one expectation for each of the ten possible actions. The expectation \IB{x{=}5} ``says'' that she will receive \$1, i.e.\ \$\IB{5{=}5}, if she guesses ``5'' and in fact $x{=}5$; but if $x{=}4$ she will receive \$\IB{4{=}5}, that is \$0.  The ``\,\MAX\,'' symbol separating the expectations is to suggest that her potential gain (i.e.\ if she acts rationally) is the maximum of the individual expected gains. If she applies that gain expression to a uniform prior over $0,\ldots,9$, the overall expected gain will be $\NF{1}{10}\Max\NF{1}{10}\Max\cdots\Max\NF{1}{10}\Max\NF{1}{10}$, i.e.\ 10\Cents\ as we saw. %
\footnote{This means that $[A \lor B]$ is not the same as $[A]\MAX\ [B]$ --- in the first case the adversary guesses whether the state is $A\lor B$, and in the second she has to choose which. For instance if her uncertainty about the secret  is modelled as uniform choice over $A,B$ then if she only ha to guess \emph{whether} the state is $A$ or $B$ then she will always be right. But if she has to \emph{choose} on or the other, then she will only be right 50\% of the time. Thus when interpreted over a distribution, these expressions yield different answers.}

But \emph{after} the \Print\ statement has leaked whether $x$ was a multiple of 3, she will have one of two beliefs: with probability \NF{1}{4} she will believe that $x$ is uniformly distributed over $0,3,6,9$; and with probability \NF{6}{10} she will believe it is uniformly distributed over $1,2,4,5,7,8$. In the two cases, the gain expression \Eqn{e1024} evaluates respectively to
{\small\begin{eqnarray}
	0\Max0\Max\NF{1}{4}\Max0\Max0\Max\NF{1}{4}\Max0\Max0\Max\NF{1}{4}\Max0 &=& \NF{1}{4} \Label{e1101-3}\\
	\NF{1}{6}\Max\NF{1}{6}\Max0\Max\NF{1}{6}\Max\NF{1}{6}\Max0\Max\NF{1}{6}\Max\NF{1}{6}\Max0 &=& \NF{1}{6} \Label{e1101-7}
\end{eqnarray}}%
and --as we saw before-- her overall expected gain from guessing \emph{after} she has run the program is \(\NF{4}{10}{\times}\EqnSmall{e1101-3}+\NF{6}{10}{\times}\EqnSmall{e1101-7} = 20\Cents\).

The gain expression \EQn{e1024} in the quantified style would be written instead \GFQ{n}{0{\leq}n{<}10}{\IB{x{=}n}}, and its evaluation over a state-distribution $\delta$ on $x$ would be
\begin{align}\Label{e1526}
	\QQ{\Max}{n}{0{\leq}n{<}10}{\EVE{x}{\delta}{\IB{x{=}n}}}\Q,
\end{align}
where now we are using syntax for generalised quantification, together with the syntax for expected value introduced in \Sec{s1715}.\,%
\footnote{The quantification is always between parentheses; and then its components come in the order
\emph{quantifier}, then \emph{bound variable(s)}; then a \emph{separator} ``$\mid$''; then any \emph{constraints} on the values the bound variables may take; then another \emph{separator} ``$\cdot$''; and finally an expression giving, for each (allowed) values of the bound variables, the \emph{expression} that participates in the quantification. Typing of the bound variables (if given) is indicated by a colon. Missing components  take default values: \True\ for a missing constraints, the tuple of bound variables for a missing expression.
\cite{Backhouse:2006vh}.}
For a ``small'' number of actions, the enumerated style \Eqn{e1024} is clearer: it is easier to see what is going on. For a ``large'' number of actions, or where they have a regular structure, the quantified style \Eqn{e1526} is sometimes better.

\Section{Reasoning algebraically \\ with gain expressions\Label{s1452}}

As we have seen just above, we use gain expressions to describe the intent of a potential adversary to explore the impact of any information leaks. Our aim is to obtain annotations of this form:
\[
\{\Exp[PreGE]\}~P~\{\,\QQ{\MAX}{i}{i\in {\cal I}}{e_i}\,\}
\]
where the post expression $\QQ{\MAX}{i}{i\in {\cal I}}{e_i}$ interpreted over the final output (hyper-distribution) of the program $P$ will give the same as the pre-expression $\Exp[PreGE]$, interpreted over the prior. (Thus providing information about the vulnerabilities on the input.)

As explained below, the the basis of verification will be to manipulate these expressions in the style of weakest pre-condition reasoning. 
Simplifications along the way can be facilitated by a \emph{gain expression algebra} for three operators  \MAX\, \PLUS\ and \AND\,,  summarised in \Fig{f0944}.

The operator \MAX\ described above represents a choice between actions -- which one the adversary chooses is on the basis of her current uncertainty of the actual value of the state, which in turn is modelled as a probability distribution. The algebra of \MAX\ in \Fig{f0944} shows it to be associative and commutative. Gain expressions also satisfy a partial order (fully explored in \cite{McIver:15}). A special case of expressions is where the adversary has no choices to pick from.

\begin{Definition}\Label{d1445}
We say that a gain expression is \emph{standard} if it corresponds to single choice for the adversary.
\end{Definition}

Next the expression $E \PLUS E'$  allows the adversary to pick independently any choice from $E$, and any choice from $E'$ and sum them together for her gain. \Fig{f0944} shows that the operator \PLUS\ is associative and commutative, and distributes through \MAX, but that \MAX\ only  weakly distributes through \PLUS.

\begin{figure}
{\small 

{\bf Properties of \MAX}
%\MAX\ is associative and commutative; it has a unit \Zero:

\begin{enumerate}[i]
\item {\bf Associative} $e_1 \MAX (e_2 \MAX e_3) = (e_1 \MAX e_2) \MAX e_3$
\item {\bf Commutative} $e_1 \MAX e_2 = e_2 \MAX e_1$
\item {\bf Unit} $e \MAX \Zero = e$
\item{\bf Dominance} If $e_1 \leq e_2$ then $e_1 \MAX e_2 = e_2$
\end{enumerate}

{\bf Properties of  \PLUS}
\begin{enumerate}[i]
\item {\bf Associative} $e_1 \PLUS (e_2 \PLUS e_3) = (e_1 \PLUS e_2) \PLUS e_3$
\item {\bf Commutative} $e_1 \PLUS e_2 = e_2 \PLUS e_1$
\item {\bf Unit} $e \PLUS \Zero = e$
\item{\bf Monotone} If $e_1 \leq e'$ then $e_1 \PLUS  e_2 \leq e' \PLUS e_2$
\end{enumerate}

{\bf Properties of \MAX\ and \PLUS\ together}
\begin{enumerate}[i]
\item {\bf Distribution} $e_1 \PLUS (e_2 \MAX e_3) = (e_1 \PLUS e_2) \MAX (e_1\PLUS e_3)$
\item {\bf Sub-distribution} $e_1 \MAX (e_2 \PLUS e_3) \leq (e_1 \MAX e_2) \PLUS (e_1\MAX e_3)$
\end{enumerate}

{\bf Properties of \MAX\, \PLUS\ and \AND\ together}
\begin{enumerate}[i]
\item {\bf Distribution-\MAX}~~ $e \AND (e_1 \MAX e_2) = (e \AND e_1) \MAX (e\AND e_2)$
\item {\bf Distribution-\PLUS}~~$e \AND (e_1 \PLUS e_2) = (e \AND e_1) \PLUS (e\AND e_2)$
\end{enumerate}

The algebra of \MAX\ and \AND\ and \PLUS\ is derived from the underlying arithmetic operators on the expectation operator and the reals.  We use $\leq$ between gain expressions as in $E \leq E'$  to mean that whenever $E$ is interpreted over a distribution $\delta$ its value is always no more than when $E'$ is interpreted over $\delta$.
}
\caption{Algebraic properties of \QIF-operators \MAX\ and \AND\ and \PLUS\Label{f0944}}
\end{figure}

Our final operator \AND\ is heterogenous, taking a standard expression as one of its arguments and a general gain expression for the other. (Both 
 \MAX\ and \PLUS\ are homogeneous.) We often use \AND\ to incorporate context where the adversary is able to be certain about relationships between program variables expressed as a boolean expression. In this case \PLUS\ behaves essentially like multiplication, distributing through \PLUS\ and \MAX.

%A standard gain expression corresponds exactly to a linear sum of the form $a\times[x{=}2]{+}b\times[x{=}3]$ \cite{McIver:05a}. 

Observe that using the algebraic rules set out at \Fig{f0944} we deduce that gain expressions similarly have a ``normal form'' based on a \MAX\ over standard expressions. 

\begin{Lemma}\Label{l1435}
Any gain expression formed using the operators \MAX\, \AND\ and \PLUS\  can be expressed as a normal form $\{ \GFQ{i}{i \in {\cal I}}{e_i} \}$, where $e_i$ are standard expressions and ${\cal I}$ is some index set.
\end{Lemma}

\LEM{l1435} tells us that although an expression can become complex, it can always be treated as a choice between some standard set of expressions, and therefore corresponds exactly to \QIF's gain expressions.

\Subsection{Reasoning with gain- and standard expressions}

Consider the following code fragment with its post gain expression \QQ{\,\MAX}{i}{0{\leq}i{<}4}{\IB{x{=}i\,}} --- this means that we are interested in 
analysing an adversary who is trying to guess the final value of $x$, and who is considering 4 possible values:

\vspace{0.5em}

\begin{tabular}{ll}
%$\{\, \Exp[Pre]\, \}$ \\[1ex]
\Assign{x}{(x \max 0) \min 2};  \quad%\\[1ex]
\Print\ \ \Exp[x\Mod 2]  \\[1ex]
$\{\,\QQ{\,\MAX}{i}{0{\leq}i{<}4}{\IB{x{=}i}\,}\,\}$ & \Cmt{Guess $x$.}
\end{tabular}

\vspace{0.5em}

An adversary who knows the program code could though, use standard  assertion-based reasoning to reason that the final value of $x$ must lies between 0 and 2, which means she would never guess that the final value of $x$ would be 3. 
How can we combine that standard assertions-based reasoning with the more intricate reasoning involving gain expressions? We use the rule above at \Fig{f1340} which allows us to use \AND\ to introduce boolean-contexts into the gain-expressions --- in this case we can say that
the expected gain under $\QQ{\MAX}{i}{0{\leq}i{<}4}{\IB{x{=}i}}$ \emph{assuming} $x\leq 2$ is equivalent to   $[x\leq 2]\AND\QQ{\MAX}{i}{0{\leq}i{<}4}{\IB{x{=}i}}$  which is \emph{the same} as the expected gain under $\QQ{\MAX}{i}{0{\leq}i{<}3}{\IB{x{=}i}}$.

\begin{figure}
\[
\begin{array}{l}
\{\Exp[PreE]\}~P~\{\Exp[E]\} ~~ \textit{and}~~ \{\Exp[PreGE]\}~P~\{\,\QQ{\MAX}{i}{i\in {\cal I}}{e_i}\,\}\\\\
\equiv~~  \{ \Exp[PreE]\AND\Exp[PreGE]\}~P~\{\,\QQ{\MAX}{i}{i\in {\cal I}}{\Exp[E]\AND e_i}\,\}
\end{array}
\]
Standard assertion-based reasoning means that the adversary can do it too. We can incorporate such assertions to simplify the gain expressions during backwards analysis.
\caption{Gain expression reasoning and standard assertions}\Label{f1340}
\end{figure} 

\Section{Reasoning backwards}% \\ with gain expressions\Label{s1258}}
\Subsection{\underline{Wh}y do we reason backwards with gain expressions?\Label{s1717}}
Backwards reasoning takes a gain expression that the adversary will apply \emph{after} a program is run (and observed), and produces a gain expression that can be evaluated over the prior distribution, i.e.\ \emph{before} running the program --- and that will give the greatest gain an adversary could realise with her chosen gain expression, the given program and that prior.\,%
\footnote{Recall \Sec{s1718}. This is motivated --inspired-- by the technique of taking a conventional post-condition, or post-expectation, and a program, and then ``working backwards'' to determine the weakest resp.\ greatest precondition or -expectation that, evaluated over the initial state resp.\ distribution, guarantees that the postcondition resp.\ -expectation will be met.}
In doing so, we --as defenders-- and she --as attacker-- are finding out \emph{in advance} how much the adversary might gain if she were to have access to our program text, the prior distribution on its state space, and the observations (leaks) that our program produces as it runs. If that value is too high (a subjective judgement), we must consider changing the program before we allow it to be run; if it is too low, she might decide to attack someone else's program instead of ours.

\Subsection{\underline{How} do we reason backwards with gain expressions?\Label{s1121}}
Here we will look at just two of the program constructions from \Fig{f1341}: assignments and conditionals.\,%
\footnote{The explicitly leaking \Print\ statement is treated in \Sec{s1626}; loops are treated in \Sec{s1309}.}
Note that the reasoning is at the source level.
\Subsubsection{Reasoning backwards with assignment\Label{s1123}}
Recall that he weakest-precondition of an assignment statement \Assign{x}{\Exp} with respect to a Boolean postcondition \Exp[post] is \Subst{\Exp[post]}{x}{\Exp}, that is \Exp[post] with all free occurrences of $x$ replaced by \Exp\ and with suitable renaming if necessary \cite{Hoare:69,Dijkstra:76}. For probabilistic programs it is the same, i.e.\ that the greatest pre-expectation of an assignment statement \Assign{x}{\Exp} with respect to a real-valued post-expectation \Exp[post] is \Subst{\Exp[post]}{x}{\Exp}, that is \Exp[post] with all free occurrences of $x$ replaced by \Exp\ \cite{Kozen:83,McIver:05a}.

With gain expressions, this same pattern is repeated: the ``greatest pre-gain'' of an assignment \Assign{x}{\Exp} with respect to an adversary's gain expression\quad\(\Exp[post1] \MAX\ \Exp[post2]\)\quad --taking a two-expectation gain expression as an example-- is
\[
	\Subst{\Exp[post1]}{x}{\Exp} \Wide{\MAX} \Subst{\Exp[post2]}{x}{\Exp}\Q,
\]
that is with the normal expectation-style substitution being applied to each \MAX\,-component of the gain expression separately. If the gain expression is written in the quantified style, the substitutions occur within the \MAX\,-quantifier. Note that assignments \emph{do not leak} from within their right-hand sides.

We illustrate this by returning to our state-space of \Sec{s1722}, but with an assignment statement as shown in \Fig{f1144}: the adversary is trying to guess the final value of $x$, and applying the rule above gives us the greatest pre-gain shown. If we were to write that out in the enumerated style, longer but easier to understand at this stage, it would be
\begin{equation}\Label{e1225}
    \begin{array}{rl}
    		 & \IB{(x\Div2)=0} ~\MAX~ \IB{(x\Div2)=1} \\
    	\MAX~& \IB{(x\Div2)=2} ~\MAX~ \IB{(x\Div2)=3} \\
    	\MAX~& \IB{(x\Div2)=4} ~\MAX~ \IB{(x\Div2)=5} \\
    	\MAX~& \IB{(x\Div2)=6} ~\MAX~ \IB{(x\Div2)=7} \\
    	\MAX~& \IB{(x\Div2)=8} ~\MAX~ \IB{(x\Div2)=9} \Q.
    \end{array}	
\end{equation}
Given that the type of $x$ is known, that simplifies to
\begin{equation}\Label{e1226}
	\begin{array}{rl}
    		&			\IB{x{=}0 \lor x{=}1}
    		    \MAX   	\IB{x{=}2 \lor x{=}3}
    			\MAX	\IB{x{=}4 \lor x{=}5} \\
    	\MAX\quad
    		&			\IB{x{=}6 \lor x{=}7}
    			\MAX	\IB{x{=}8 \lor x{=}9} \Q,
	\end{array}
\end{equation}
since the expectations in \Eqn{e1225} from \IB{(x\,\Div\,2)=5} onwards are 0 over all possible distributions on $\{0,\ldots,9\}$. In the quantified style \Eqn{e1226} is more compactly
\begin{equation}\Label{e1257}
	\GFQ{d}{0{\leq}d{<}5}{\IB{x{=}2d\lor x{=}2d{+}1}}\Q.
\end{equation}

\begin{figure}
\begin{tabular}{ll}
$\{\,\QQ{\MAX}{n}{0{\leq}n{<}10}{\IB{(x\Div2)=n}}\,\}$ \\[1ex]
\Assign{x}{x\Div2} \\[1ex]
$\{\,\QQ{\MAX}{n}{0{\leq}n{<}10}{\IB{x{=}n}}\,\}$ & \Cmt{Guess $n$.}
\end{tabular}

\caption{Guessing a number between 0 and 9\Label{f1144}}
\end{figure}

No matter how the pre-gain is written, e.g.\ whether as in \Fig{f1144}, or as at (\ref{e1225}, \ref{e1226} or \ref{e1257}), knowing the prior distribution $\pi$ of $x$ we can use it to determine \emph{beforehand} how much the adversary can gain by trying to guess $x$ \emph{after} \Prog{f1144} was run. If for example the prior $\pi^3$ were that $x$ is uniformly distributed over multiples of 3 only (i.e.\ 0,3,6,9 with probability \NF{1}{4} for each), then using \Eqn{e1226} we can see that her expected gain is
\[ \NF{1}{4} ~\Max~ \NF{1}{4} ~\Max~ 0 ~\Max~ \NF{1}{4} ~\Max~ \NF{1}{4}\Q, \]
that is \NF{1}{4}, because for example \EVE{x}{\pi^3}{\IB{x{=}0 \lor x{=}1}} is \NF{1}{4}, but \EVE{x}{\pi^3}{\IB{x{=}4 \lor x{=}5}} is 0.

\Subsubsection{Reasoning backwards through \If-conditionals\Label{s1124}}
As for assignments, our treatment of conditionals will be based on the extant techniques for \Kuifjetje's antecedents. Given a (post-) gain-expression \Exp[post] and (e.g.\ \Fig{f1341}) a conditional 
\[ \If\ \Exp[cond\,]~~\Then\ \Exp[stmtT]~~\Else\ \Exp[stmtE] \Q,\]
the first step is to determine separately the pre-gains for the bodies \Exp[stmtT] and \Exp[stmtE] with respect to that \Exp[post]. Call them \Exp[preT] and \Exp[preE] respectively. If the original \Exp[post] used the operator \MAX\,, as it would have if there were more than one expectation there, then it is likely that \Exp[preT] and \Exp[preE] will contain \MAX\,'s as well. (If they don't, it will likely be because they have been simplified away.)

The second step is to take the condition \Exp[cond] into account: the \Then\ pre-gain \Exp[preT] becomes \IB{\Exp[cond]}\,\AND\,\Exp[preT], where the operator \AND\ takes a non-negative real scalar $c$ say on the left and a gain expression on the right, and multiplies the scalar into each expectation of the gain expression. Thus for example
\[
	c ~\AND~ (\Exp[exp1] \MAX\, \Exp[exp2])
	\Wide{=}
	c{\times}\,\Exp[exp1] ~\MAX~ c{\times}\,\Exp[exp2] \Q.
\]
For the \Else\ branch, the same applies except that the scalar $c$ is \IB{\neg\Exp[cond]}.

If there is no \Else\ branch, it is taken to be \Else\ \Skip, in which case we have $\IB{\neg\Exp[cond]} \AND\, \Exp[post]$, because \Skip\ acts as the identity on its post-gain.

The third (and final) step is to add the two scalar-multiplied pre-gains together, the \Then- and the \Else-, using a ``lifted'' addition operator that must be distributed into any \MAX\,'s that occur in its arguments.\,%
Thus for example
\[
	\begin{array}{ccccc}
		& \multicolumn{4}{l}{(\Exp[preT1]\MAX\Exp[preT2]) ~\PLUS~ (\Exp[preE1]\MAX\Exp[preE2])} \\[1ex]
	=	&		& (\Exp[preT1]+\Exp[preE1]) 
		&\MAX	& (\Exp[preT1]+\Exp[preE2]) \\
		&\MAX	& (\Exp[preT2]+\Exp[preE1]) 
		&\MAX	& (\Exp[preT2]+\Exp[preE2]) \Q,
	\end{array}
\]
where the $+$'s occuring in the expectations (on the right) are ordinary addition of expressions.

\FIG{f1544} gives an example --- and it is (unexpectedly) subtle. An \emph{informal} analysis might be as follows:
\begin{figure}
\begin{tabular}{lr}
$\{\,\IB{a\lor b} ~~\MAX~~ \IB{a\lor \neg b}\,\}$ & \Cmt{\emph{Really?}}\\[1ex]
%\NoDo{$\{\,\IB{a}+\IB{\neg a\land b} ~~\MAX~~ \IB{a}+\IB{\neg a\land \neg b}\,\}$} & \Cmt{\emph{Really?}}\\[1ex]
\If\ $a$ \Then\ \Assign{c}{\True} \Else\ \Assign{c}{b} \\[1ex]
$\{\,\IB{c} \MAX \IB{\neg c}\,\}$ & \Cmt{Guess $c$.}
\end{tabular}

\bigskip
The state-space comprises three Booleans $a,b,c$ and, once the program has run, the adversary will guess the value of $c$. The program contains no \Print\ statement, i.e.\ has no explicit leak.

\smallskip\quad 
Functionally, the program is equivalent to \Assign{c}{a\lor b}. Why then doesn't \Sec{s1123} give \IB{a\lor b} \MAX\,\,\IB{\neg a\land \neg b} for the pre-gain, agreeing with \Eqn{e1949}?
\caption{Guessing a Boolean\Label{f1544}}
\end{figure}
\footnotetext{If $a,b$ are certainly \False\ (i.e.\ $\alpha{=}\beta{=}0$) then of course she will guess \False\ for $c$. And if either of $a,b$ is certainly \True\ ($\alpha{=}1$ or $\beta{=}1$) then she will guess \True\ for $c$. But e.g.\ when $\alpha=\beta=1{-}\NF{1}{\sqrt2}\sim 0.3$, her reasoning above would not give a clear favourite guess for $c$.} 

Suppose for simplicity that the prior probabilities of $a,b$\,'s being \True\ are independently $\alpha,\beta$ respectively. Because functionally the program has the effect \Assign{c}{a\lor b}, the adversary could calculate that $c$ is therefore \False\ with probability \Not{\alpha}\Not{\beta} (using \Not{\alpha} for $1{-}\alpha$ etc.)\ and (therefore) \True\ with probabilty $1-\Not{\alpha}\Not{\beta}$. Her rational guess for $c$ would be to choose the Boolean for which the corresponding probability is greater, and in that case she would be right with probability
\begin{equation}\Label{e1949}
	1{-}\Not{\alpha}\Not{\beta} \Wide{\Max} \Not{\alpha}\Not{\beta} \Q.\hspace{2em}
\end{equation}
That is supported by the pre-gain \IB{a\lor b} \MAX\,\,\IB{\neg a\land \neg b} mentioned in the text of \Fig{f1544}. \emph{But it is incorrect.}

The \emph{actual} pre-gain takes into account that in \Kuifjetje
\begin{quote}
	\emph{Conditional statements leak the test unconditionally},
	i.e.\ irrespective of the program code in the two branches (one of which might be \Skip).
\end{quote}
``Branching on high'' is not to be done lightly, and we call that an ``implicit'' leak (as opposed to the explicit leak of a \Print). The reason for the implicit leak is discussed below (\Sec{s1357}); but for now we assume it.

That being so, the \Kuifjetje-correct informal reasoning is as follows, based on the same prior as in \Fig{f1544}. With probability $\alpha$, Boolean $a$ will be \True\ and the adversary will know that (because of the implicit \If-leak), and she will know further from the program code that $c$ will be set to \True. If $a$ is \False\ (probability \Not{\alpha}), then she will know that $c{=}b$ finally, and so the most likely value for $c$ will be the same as the most likely value for $b$. Her probability of being right is in that case $\beta\Max\Not{\beta}$.

And so her probability \emph{overall} of guessing $c$ correctly is $\alpha{\times}1+\Not{\alpha}{\times}(\beta\Max\Not{\beta})$, simplified\ $\alpha+\Not{\alpha}(\beta\Max\Not{\beta})$, which is indeed what the pre-gain in \Fig{f1544}, viz.
%\begin{equation}\Label{e1407}
% 	\NoDo{\IB{a}+\IB{\neg a\land b} ~~\MAX~~ \IB{a}+\IB{\neg a\land \neg b}}
%\end{equation}
\begin{equation}\Label{e1407}
 	\IB{a\lor b} ~~\MAX~~ \IB{a\lor \neg b} \Q,
\end{equation}
gives if evaluated over the prior we chose for our example.

%\ToDo{Check that the red versions are correct.}

The detailed source-level reasoning for conditionals, explained above, works out as follows for \Fig{f1544}. We start with the post-gain \IB{c} \MAX\, \IB{\neg c} and work backwards.
\begin{description}
\item[\small\emph{First step}, \Then\ branch is \Assign{c}{\True}\,.]~\\
Pre-gain is \IB{\True}\,{\MAX}\,\IB{\False}, which simplifies to 1.
\item[\small\emph{First step}, \Else\ branch is \Assign{c}{b}\,.]~\\
Pre-gain is \IB{b}\,\MAX\,\IB{\neg b}.\,
\footnote{Note that \IB{b}\,{\MAX}\,\IB{\neg b} does \emph{not} simplify to just 1, which is what case analysis over ``$b$ is \True\ or \False'' might suggest. For \IB{b}\,{\MAX}\,\IB{\neg b} is a gain-expression with \emph{two} expectations (i.e.\ two potential adversary actions), whereas 1 is a gain-expression with just a single action (and the adversary has no choice). If in distribution $\delta$ the Boolean $b$ were \True\ with probability $\beta$, then the former gain-expression would evaluate to $\beta\Max\Not{\beta}$, i.e.\ it depends on $\delta$. The latter would evaluate to 1 for all distributions $\delta$.} 
\item[\small\emph{Second step}, \Then-condition is $a$\,.]~\\
Conditioned \Then-pre-gain is $\IB{a}\,{\AND}\,1 = \IB{a}$.
\item[\small\emph{Second step}, \Else-condition is $\neg a$\,.]~\\
Conditioned \Else-pre-gain is $\IB{\neg a}\AND(\IB{b}\,{\MAX}\,\IB{\neg b})$ which, after \AND\,-distribution, becomes $\IB{\neg a\land b}\,\MAX\,\IB{\neg a\land \neg b}$.
\item[\small\emph{Third step}, summation of the branches' conditioned pre-gains.]~\\
The sum is $\IB{a} ~\PLUS~ (\IB{\neg a\land b}\,\MAX\,\IB{\neg a\land \neg b})$ which, when the \PLUS\ is distributed-in and the Booleans simplified, gives \Eqn{e1407} as above (agreeing with ``\emph{Really?}''\ in \Fig{f1544}.)
\end{description}

\Subsection{Reasoning about \Print\ and \While\ statements}
These are deferred to Secs. \ref{s1626} and \ref{s1309} respectively.

\Section{The attack model\Label{s1627}}
\Subsection{What the adversary knows in all cases\Label{s1044}}
In \Sec{s1637}(\ref{i1405}) we mentioned the first of the following:\,%
\begin{enumerate}
\item\Label{i1044-1} All \Kuifjetje\ variables are hidden from the adversary, i.e.\ they are all ``high security''. There are no ``visible'', or ``low security'' variables per se. (But see \Sec{s1626a} below.)
\item The state-space prior \emph{distribution} is known to the adversary, but the precise initial state \emph{itself} is not.
\item The program code is known to the adversary. (No ``security through obscurity'' \cite{Kerckhoffs:1883tm,Alvim:20a}.)
\item\Label{i1044-4}The program-counter is known --or can be inferred-- by the adversary as the program runs \cite[\S13.4.3]{Alvim:20a}.
%\ToDo{Do we need more justification for this \Eqn{i1044-4}? Maybe \App{a1256}.}
\end{enumerate}
Those are conventional (though perhaps not universal) attack-model assumptions.

\Subsection{Explicit leaks in the attack model\Label{s1249}}
Explicit leaks in our attack model are represented by \Printje\ statements that express, as a program-variable expression, an escape of information at that point that we know is unavoidable given the purpose of the program and the constraints on how it is implemented (e.g.\ the specific hardware). For example, if portions of the program's final state are provided to the visible to everyone, there should be a \Printje\ of them at the program's end (as in \Fig{f1544} where there should be a \Print\ $L$). But further, if e.g.\  a particularly tight loop at some point in the program's code generates measurable heat, then a\quad \Print\ ``hot''\quad might be put in the code at that point.

One consequence of \Eqn{i1044-4} in particular is that statements $\Printje\ a$ and $\Printje\ \neg a$ are equivalent because, knowing the program code and the program counter, the adversary will deduce the value of the (hidden) variable $a$ either way: since she knows what the \Printje-expression was, she will know whether to negate what she observed. 

Finally, explicit (though harmless) information leaks are leaks of constants, or even leaks of values that can be deduced by the adversary with conventional assertional reasoning: for example both \Printje\ \True\ and \Printje\ \False\ are equivalent to \Skip; and \Printje\ $x$ \emph{following} \Assign{x}{1} is as well: in all three cases, the adversary knows what is leaked simply by looking at the program code --- she has no need to see it happen.

In summary: explicit leaks are caused by (ordinary) assertional reasoning, by \Printje\ statements or (\Sec{s1626a} below) by the use of ``visible'' variables.

\Subsection{Implicit leaks\Label{s1357}}
An \emph{implicit} information leak occurs when the program counter reveals which branch of a conditional choice was taken: she learns whether its condition evaluated to \True\ or \False\ at the branching moment. And since she knows the program code, she knows what Boolean expression the condition evaluated.

This implicit-leak semantics is \emph{built-in} to \Kuifjetje\ --- but not because ``branching on high'' is a bad idea, and should be discouraged. Rather it is because of referential transparency, i.e.\ to allow reliable and modular reasoning, and so its inclusion to some extent validates that principle. It is not a ``semantic bug'' --- it is a necessity.

An example of such reasoning is as follows. Having the program-equality $\Print\ \True = \Print\ \False = \Skip$, as above, we have by referential transparency also the equality
\begin{align*}
				&	\If\ a~ \Then\ \Skip\ \Else\ \Skip \\
	\WIDE{=}	&	\If\ a~ \Then\ \Print\ \True\ \Else\ \Print\ \False \Q.
\end{align*}
Since an observer of the right-hand program will deduce $a$'s value from whether she sees \True\ (from the \Printje\ \True) or \False, she must be able to deduce $a$'s value in the left-hand program as well. (Similar appeals to referential transparency occur in \cite{Alvim:20a,Morgan:07}.)

\Subsection{\Print\ statements\Label{s1626}}
Putting the above together gives us the source-level rule for reasoning about \Print\ statements (but we will for the moment consider only the \Print'ing of Booleans). The statement \Print\ \Exp[bexp]\ is ``sugar'' for the trivial\quad\If\ \Exp[bexp]\ \Then\ \Skip\,,\quad whence we can simply apply reasoning for conditionals. (\Fig{f1215} gives an example.)
Thus --referring to \Sec{s1124}-- we have have \Fig{f1456}, this time by calculation rather than intuition (as it was in \Fig{f1645}).
\begin{figure}
\begin{tabular}{l}
\{\,~\hspace{1.95em}\GfQ{\,n{\In}\{0,3,6,9\}}{\hspace{1.85em}\IB{x{=}n}} \\
~~~\PLUS~ \GfQ{\,n{\In}\{1,2,4,5,7,8\}}{\IB{x{=}n}}\,\} \\[1ex]
\Print~~$(x\!\!\mod3){=}0$ \\[1ex]
$\{ \GFQ{n}{0{\leq}n{<}10}{\IB{x{=}n}} \}$
\end{tabular}

\bigskip
\caption{Guessing a number $x$ where $0{\leq}x{<}10$\Label{f1456}}
\end{figure}
The first (\,\If\,) step is trivial, because both branches are \Skip. The second and third steps together start with \begin{eqnarray*}
						&& \IB{(x\Mod3)=0}    \,\AND\, \GFQ{n}{0{\leq}n{<}10}{\IB{x{=}n}} \\
		&\Wide{\PLUS}	&  \IB{(x\Mod3)\neq0} \,\AND\, \GFQ{n}{0{\leq}n{<}10}{\IB{x{=}n}} \\[1ex]
	=					&& \GFQ{n}{0{\leq}n{<}10}{\IB{(x\Mod3){=}0 \land x{=}n}} \\
		&\Wide{\PLUS}	&  \GFQ{n}{0{\leq}n{<}10}{\IB{(x\Mod3){\neq}0 \land x{=}n}} \\[1ex]
	=					&& \GFQ{n}{n{\in}    \{0,3,6,9\}}{\IB{x{=}n}} \\
		&\Wide{\PLUS}	&  \GFQ{n}{n{\in}\{1,2,4,5,7,8\}}{\IB{x{=}n}} \\[1ex]
	=					&& \GfQ{n{\In}    \{0,3,6,9\}}{\IB{x{=}n}} \\
		&\Wide{\PLUS}	&  \GfQ{n{\In}\{1,2,4,5,7,8\}}{\IB{x{=}n}} \Q{which,}
\end{eqnarray*}
simplified, leads to the expression just above for the pre-gain (but where we have not yet distributed the \PLUS\ into the \MAX\,'s). For the uniform distribution on $x$, that (pre-) gain-expression evaluates to $\NF{1}{10}{+}\NF{1}{10} = 20\Cents$, as earlier just before \Eqn{e1526}.

\Subsection{Visible (low level) vs.\ hidden (high level) variables\Label{s1626a}}
At \Sec{s1044}\Eqn{i1044-1} we stated that all \Kuifjetje\ variables are hidden: this simplifies the semantics drastically. But in conventional presentations, it's very convenient to give variables an attribute of ``low'' or ``high'', so that low variables are visible (to everyone, including the adversary), while high variables remain hidden.
Thus when we want to treat some variables as always visible, we introduce the sugar that any variable $v$ given the attribute ``visible'' has an invisible ``$\Printje\ v$'' inserted immediately after any assignment to it \cite[\S15.2]{Alvim:20a}.
%\footnote{We won't however use the ``visible'' attribute in our examples here.}
For example in \Eqn{e1405} we could declare $L$ to be \emph{visible}, in which case that program would actually be
\begin{quote}\begin{tabular}{l}
	\If\ $\;H\Mod 8 = 0\;$ \\
		\quad\Then\ \Assign{L}{H}; \Print\ $L$ \\
		\quad\Else\ \Assign{L}{1}; \Print\ $L$
	\end{tabular}
\end{quote}
\FIG{f1123} gives a further example of visible variables.

\Section{Reasoning backwards through \While-loops\Label{s1309}\Label{a1052}}

%{\color{blue}%1152
As for \If-conditionals, for \While-loops we start with a post-gain \Exp[post], and work backwards towards the loop's pre-gain (which will be its \QIF-invariant for that \Exp[post]).

As for \If, there are three calculations for determining a \While\ pre-gain given
\begin{quote}
\quad\While\ G\ \Do\ \Exp[body]\ \Od\ \{\,\Exp[post]\,\}
\end{quote}
They are to show that:
\begin{eqnarray}
	\Exp[pre]      &\Eqv& \Exp[thenPre]\; \PLUS\ \Exp[elsePre] \Label{e1052a}\\
	\Exp[thenPre]  &\Eqv& \IB{G}\ \AND\ \GPG{\Exp[body]}{\Exp[pre]} \Label{e1052b}\\
	\Exp[elsePre] &\Eqv& \IB{\neg G}\ \AND\ \Exp[post] \Q, \Label{e1052c}
\end{eqnarray}
where\quad\GPG{\Exp[body]}{\Exp[pre]}\quad is the \emph{greatest pre-gain} determined by applying our rules (structurally recursively) to \Exp[body] with post-gain \Exp[pre].\,%
\footnote{The \If- and the \While-rules differ in the post-gains of the second- and third steps, in particular that the second \While-step ``recursively'' uses \Exp[pre].}
%\ToDo{Think about how much we should talk about ``greatest''.}

Below is how it works out for our example in \Fig{f1356}, where $A[{:}n]$ means ``up to but not including $A[n]$'', and $A[n{:}]$ means ``from $A[n]$ to the end''. We take \While-\Exp[pre] as \IB{\,x{\in}A[n{:}]\,}, and \Fig{f1355} supplies ``standard'' assertional reasoning (for the same program) where required: when we use \Fig{f1355}, there will be a Boolean ``justification'' at right,
%\footnote{\ToDo{Do we really need that appeal?}}
except that the standard (Boolean) invariant $0{\leq}n{\leq}N$ is used silently.

\begin{figure}
\begin{tabular}{lr}
$\{\,0\,{\leq}\,N\,\}$ &\Cmt{Conventional precondition}\\
$n$:= $0$ \\
\While\ $n{\neq}N \land A[n]{\neq}x$ \Do\ &\Cmt{\Exp[Inv]: $0{\leq}n{\leq}N \land x{\not\in}A[{:}n]$}\\
\quad $n$:= $n+1$ \\
\Od \\
$\{\,$ \Exp[Inv] $\land\; n{=}N\,\}$ \\
$\{\,x{\not\in}A\;\text{if}\,n{=}N\,\text{else}\;A[n]{=}x\,\}$ &\Cmt{Conventional postcondition}
\end{tabular}

\caption{Looking for $x$ in an array $A$ of size $N$: conventional presentation\Label{f1355}}
\end{figure}

\begin{figure}
\begin{tabular}{lr}
$\{\,\IB{x{\in}A}\,\}$ &\Cmt{Via assignment rule.}\\
\Assign{n}{0} \\[0.5ex]
\{\,\IB{\,x{\in}A[n{:}]\,}\,\} & \Cmt{\While\ Pre-gain} \\
\While\ $n{\neq}N \land A[n]{\neq}x$ \Do \\
\quad $n$:= $n+1$ \\
\Od \\
\GFQ{i}{0{\leq}i{<}N}{\IB{A[i]{=}x}} &\Cmt{Post-gain}
\end{tabular}

\bigskip
The \While's \Exp[post] expresses that the adversary's action is to choose any $0{\leq}i{\leq}N$ and then guess that $A[i]{=}x$, gaining \$1 if she chooses rationally.

And because of the implicit leak of $n$ induced by the \While-loop guard, she is guaranteed to gain that \$1 \emph{unless} $A$ does not contain $x$ at all,
in which case her guess will be wrong no matter whichever $i$ she chooses, thus gaining \$0.
Those two outcomes are expressed by the \While's \Exp[pre] being \IB{\,x{\in}A\,}.

\caption{Looking for $x$ in an array $A$ of size $N$: \QIF\ presentation\Label{f1356}}
\end{figure}

%I'll use \Eqv\ wherever possible. 
\begin{description}
\item[]
\item[for \Eqn{e1052c},] i.e.\ \Exp[elsePre], we have that \quad $\IB{\neg G}\ \AND\ \Exp[post]$
\begin{Reason}
\Space
\Step{is}{&\IB{n{=}N\lor A[n]{=}x}\\ \AND& \GFQ{i}{0{\leq}i{<}N}{\IB{A[i]{=}x}}}
\Space
\StepR{\Eqv}{$x{\not\in}A[{:}n]$}{&\IB{n{=}N\lor A[n]{=}x}\\ \AND& \GFQ{i}{n{\leq}i{<}N}{\IB{A[i]{=}x}}}
\Space
\Step{\Eqv}{\GFQ{i}{n{\leq}i{<}N}{\IB{\,n{=}N\lor A[n]{=}x\,}\times\IB{A[i]{=}x\,}}}
\Step{\Eqv}{\GFQ{i}{n{\leq}i{<}N}{\IB{\,(n{=}N\lor A[n]{=}x)\land A[i]{=}x\,}}}
\Space
\Step{\Eqv}{\IB{\,n{\neq}N \land A[n]{=}x\,}\Q.}
\end{Reason}

\item[]
\item[for \Eqn{e1052b},] i.e.\ \Exp[thenPre], we have that \quad $\IB{G}\ \AND\ \GPG{\Exp[body]}{\Exp[post]}$
\begin{Reason}
\Space
\Step{is}{&\IB{n{\neq}N\land A[n]{\neq}x}\\ \AND& \GPG{(n\text{:=}n{+}1)}{\IB{\,x{\in}A[n{:}]\,}}}
\Space
\Step{\Eqv}{\IB{\,n{\neq}N\land A[n]{\neq}x\,} \;\AND\; \IB{\,x{\in}A[n{+}1{:}]\,}}
\Step{\Eqv}{\IB{\,n{\neq}N\land A[n]{\neq}x \land x{\in}A[n{+}1{:}]\,}\Q.}
%\Space
%\StepR{\Eqv}{$x{\not\in}A[{:}n{+}1]$}{\IB{\,n{\neq}N\land x{\in}A[n{:}]\,}\Q.}
\end{Reason}

\item[]
\item[for \Eqn{e1052a},] i.e.\ \Exp[pre], we have that \quad \Exp[thenPre]\; \PLUS\ \Exp[elsePre]

\begin{Reason}
\Step{is}{         &\IB{\,n{\neq}N\land A[n]{\neq}x \land x{\in}A[n{+}1{:}]\,} \\
           \PLUS\; &\IB{\,n{\neq}N \land A[n]{=}x\,}}
\Space
\Step{\Eqv}{\IB{\,n{\neq}N\land x{\in}A[n{:}]\,}}
\StepR{\Eqv}{$A[N{:}]$ is empty}{\IB{\, x{\in}A[n{:}]\,}}
\Step{\Eqv}{\Exp[pre]\Q,}
\end{Reason}
as required.
\end{description}

The \emph{overall} pre-gain of \Fig{f1356} is then \GPG{(\Assign{n}{0})}{\Exp[pre]}\,, that is \IB{\,x{\in}A[0{:}]\,}\,, equivalently \IB{\,x{\in}A\,}\,.
Interpreted intuitively, the gain expression \IB{\,x{\in}A\,} says that the adversary will gain \$1 if $x$ occurs \emph{anywhere} in $A$ because --if it does-- the index will have been leaked as well. But how was it leaked?

\begin{figure}
\begin{tabular}{lr}
\multicolumn{2}{l}{$\{\,\IB{\,x{\in}A\,}\,\}$\hspace{4em}\Cmt{Pre-gain: attack succeeds when $x{\in}A$}} \\[0.5ex]
\Assign{n,b}{0,\False}; \Print\ b &\Cmt{$b$ is visible} \\
%\Cmt{Inv:\quad\GFQ{i}{0{\leq}i{<}n}{\IB{\,A[i]{=}x\,}} \MAX\ \IB{\,x{\in}A[n{:}]\,}}\\
\While\ $n{\neq}N$ \Do \\
\multicolumn{2}{l}{\quad \If\ $A[n]{=}x$ \Then\ \Assign{b}{\True}; \Print\ b\hspace{4em}\Cmt{$b$ is visible}} \\
\quad $n$:= $n+1$ \\
\Od \\
\Cmt{Here $b= (x{\in}A)}$\,. \\
%\Print\ b \\
\{\,\GFQ{i}{0{\leq}i{<}N}{A[i]}\,\} &\Cmt{Post-gain (the attack)}
\end{tabular}

\bigskip
Here we treat Boolean $b$ as a ``visible'' (low security) variable, in the style of \Sec{s1626a}. But assignments to $b$ are \emph{not} the cause of the leak here: see \Fig{f1215}.
\caption{Alternative to \Fig{f1356}\Label{f1123}}
\end{figure}
Investigating that, in \Fig{f1123} we run the loop to completion every time. Whether $x$ is in $A$ is now given explicitly by Boolean $b$, which we treat as a ``visible'' (low security) variable to give an example for \Sec{s1626a}.

But --unlike \Fig{f1356}-- in this case the loop iterates $N$ times exactly. Does the adversary nevertheless \emph{still} learn where $x$ is? Yes --surprisingly-- the assignments to $b$ are not the cause of the leak: see \Fig{f1215}.

The examples of Figs.\ \ref{f1356},\ref{f1123},\ref{f1215} are the kind of analysis that could be used to study the impact of information leaks in a library for searchable encryption \cite{JuradoP021}. Such a specification would require that the information flow \emph{only} determines whether or not an item $x$ is in the database (in this simple form, the array), but should not identify the record number. Here the analysis shows that for an implementation that stops early (\Fig{f1356}) the adversary could learn the record --- such implementations therefore could be vulnerable to a timing attack. Our analysis also shows that implementation that does not stop early (\Fig{f1215})  also leaks the record however because of a ``branch on high vulnerability''. A secure implementation would have to ensure that neither of these vulnerabilities were present.

\begin{figure}
\begin{tabular}{lr}
\multicolumn{2}{l}{$\{\,\IB{\,x{\in}A\,}\,\}$\hspace{4em}\Cmt{Pre-gain: attack succeeds when $x{\in}A$}} \\[0.5ex]
%$\{\,\IB{\,x{\in}A\,}\,\}$ &\Cmt{Pre-gain} \\[0.5ex]
\Assign{n}{0} \\
\While\ $n{\neq}N$ \Do \\
\quad \If\ $A[n]{=}x$ \Then\ \Skip &\Cmt{Equivalently \Print\ $A[n]{=}x$}\\
\quad $n$:= $n+1$ \\
\Od \\
\{\,\GFQ{i}{0{\leq}i{<}N}{A[i]}\,\} &\Cmt{Post-gain (the attack)}
%\{\,\GFQ{i}{0{\leq}i{<}N}{A[i]}\,\} &\Cmt{Post-gain}
\end{tabular}

\bigskip
The leak is caused by the branch on high ``\,\If $A[n]{=}x$\,'', not by the assignments to $b$ (which variable is no longer there).
\caption{Alternative to Figs.\,\ref{f1356},\ref{f1123}\Label{f1215}}
\end{figure}

\Section{Mathematical Foundations\Label{s1059}}

In this section we summarise the mathematical foundations for reasoning about quantitative information flow in programs \cite{McIver:15,McIver:11,McIver:10}. The approach extends both \QIF\ for channels \cite{Alvim:2012aa,McIver:13} and probabilistic program semantics \cite{Kozen:83,McIver:05,Kaminski19,0001KKOGM15}. Those mathematical foundations feature the following:

%\begin{enumerate}[(i)]
%\item 
\underline{{\bf Secrets}} are possible values
%of  the state (defined by the program variables)
drawn from   $S$,  the set of states. 
The adversary's knowledge of the state is modelled as a probability distribution $\pi{\In}\Dist S$. We often assume that the input to the program (from the point of view of an adversary) is essentially a (prior) distribution. 

%\item\label{ac}
\Subsection{Accounting for information leaks in programs\Label{diamond}}
%\noindent($\diamondsuit$) \underline{Accounting for information leaks in programs}.
An {\bf  abstract channel} \cite{Alvim:20a} is the model we use for describing information leaks, such as \Printje\ statements and conditionals; it is a function of type $\Dist S\Fun\Dist^2 S$ satisfying a number of ``healthiness conditions'' (set out below at \ref{heart}). Here, $\Dist^2 S$ is a {\bf hyper-distribution} which ``abstracts'' from an observation type,  leaving precisely the structure needed to evaluate leakages, namely the marginal distribution over the observations and the conditional distributions associated with each observation. These are present in the  structure $\Dist^2 S$ as an (outer) distribution over posteriors of type $\Dist S$, where the outer distribution corresponds to the marginal over observations and the posteriors correspond to the conditional distribution for each marginal. We use hyper-distributions because they satisfy a number of mathematical properties convenient for the features of the more general program semantics we need here. % Said that as your first sentence already.
%In the programming language, conditionals and print statements are modelled as channels.
%To illustrate, 
The information leak caused by the \If-conditional in \Fig{f1544} is equivalent to \If\ $a$ \Then\ \Skip\ \Else\ \Skip\,, equivalently \Print\ $a$.
%\quad or, more succinctly, just \Print\ $a$.
 
That has a channel representation in terms of a stochastic, i.e.\ channel matrix where the rows correspond to the state (for example) determined by the variables $a$ and $b$, and the columns correspond to the observations, i.e.\ whether $a$ is true or not. Each row of the channel therefore is the probability with which a given state produces the observation of the column. In this deterministic case all the probabilities are 0 or 1:
\newcommand\AT {{\sf T}}
\newcommand\AF {{\sf F}}
\[
\begin{array}{c | cc}
(a,b) & \textit{a is \AT} & \textit{a is \AF}\\
\hline
(\AT,\AT) & 1 & 0\\
(\AT,\AF) & 1 & 0\\
(\AF,\AT) & 0 & 1\\
(\AF,\AF) & 0 & 1\\
\end{array}
\]

%\item \label{mm} 
%\noindent($\clubsuit$) \underline{Program updates}, 
\Subsection{Program updates such as assignment statements\Label{club}}
These are modelled as {\bf Markov} matrices of type $S\Fun\Dist S$, widely used  for probabilistic semantics \cite{Kozen:83,McIver:05,Kaminski19}.  

As an example, the update to the variable $c$ in  program at \Fig{f1544} is modelled as a (deterministic) Markov matrix which maps an input state $(a, b, c)$ to an output state of $(a, b, a \lor b)$. %Again, it is deterministic.

%\item\Label{infModel}
%\noindent ($\spadesuit$) 
\Subsection{Model for information-leakage in programs\Label{spade}}
\Kuifjetje\ programs are modelled as a combination of both \ref{diamond}
%$\diamondsuit$
and \ref{club} sitting overall within a model of {\bf abstract} Hidden Markov Models \cite{McIver:15} of type $\Dist S\Fun\Dist^2 S$ that combine an information leak \ref{diamond}  with an update \ref{club} (either or both of which could be trivial: no leak, or identity update). In order to instil Markov updates with the correct type we use structural embedding via the {\bf Probabilistic Monad} \cite{Jones:89,Giry:81} defined by $(\Dist, \Avg, \Unit)$. Here $\Dist$ is a type constructor which given a metric space  $(S,d)$ creates the space of probability distributions $\Dist S$ from the Borel algebra induced by $d$.  The function $\Unit{\In}S{\Fun}\Dist S$ creates a point distribution centred on its argument; and the function $\Avg\In\Hyp S\Fun\Dist S$ averages over its input hyper-distribution.%

Now, given a Markov matrix $M\In S\Fun\Dist S$ we can embed it in the space of \HMM's using the structural functions (above) of the Probabilistic Monad, which therefore  preserve its essential probabilistic features:
in this case the result should be the point hyper constructed as follows. Given an input distribution $\pi$ of type $\Dist S$, apply the push-forward $\Dist M$ of $M$ to it to give a result of type $\Dist^2 S$, and then \Avg-average that to  give $\Avg(\Dist M(\pi))$ of type $\Dist S$ again. Use \Unit\ to convert that into a point-hyper, giving $\Unit(\Avg(\Dist M(\pi)))$, so that our original Markov matrix $M$ (a function of type $S{\Fun}\Dist S$) becomes a ``point'' \HMM\ $(\Unit\Comp\Avg\Comp\Dist)(M)$ of type $\Dist S{\Fun}\Dist^2S$, now a mapping of a prior to a hyper-distribution as it should be.

{\bf Sequential composition} in the {\HMM} space is now also a structural definition: given two \HMM's $H_1, H_2$, their sequential composition $H_1 ; H_1$ is $\Avg\Comp (\Dist H_2)\Comp H_1$, that is the standard Kleisli construction. Thus $H_1$ is applied to an incoming $\pi$ resulting in a hyper-distribution (i.e. the digest of the information leaks as well as the state updates rendered by $H_1$); the push-forward of $H_2$ then follows (essentially applying $H_2$ to each posterior in $H_1(\pi)$), resulting in an output of type $\Dist^3 S$, which the final application of $\Avg$ then ``amalgamates''.

As an example, the final hyper-distribution for \Fig{f1544} can be computed using the channel matrix representation for the leak at \ref{diamond}  followed by the Markov Matrix state update given at \ref{club}. Assuming the prior described at \Fig{f1544} where $a$ is \AT\ with probability $\alpha$ and $b$ is \AT\ (independently) with probability $\beta$, the final hyper-distribution has two associated posteriors, corresponding to the two observations (i.e.\ the two possible values \AT,\AF\ leaked by the test of $a$). Because $c$ finally must be the disjunction of $a$ and $b$, there are only 4 possible final states. We summarise the hyper-distribution below, where the columns are now labelled with the marginal probabilities corresponding to the possible observations of $a$, and the contents of each column represents the conditional probabilities over the state:
\[
    \begin{array}{c|cc}
        (a,b,c) & \alpha & 1{-}\alpha\\
        \hline
        (\AT,\AT,\AT) & \beta & 0\\
        (\AT,\AF,\AT) & 1{-}\beta & 0\\
        (\AF,\AT,\AT) &  0 &\beta \\
        (\AF,\AF,\AF) &0 & 1{-}\beta \\
    \end{array}
\]
\Subsection{Healthiness conditions for \QIF\ programs\Label{heart}}
An advantage of using the probabilistic monad is that it brings with it a large number of mathematical properties, which is the reason our definitions use the monad's structures; and the basic definitions of state update and conditional are proven to be continuous arrows in the underlying category of continuous functions of measures. For example programs are all continuous wrt.\ the metric spaces defined above, sequential composition is a continuous operator between programs.

\Subsection{Measuring leakage, entropies, gain functions}
A {\bf vulnerability} is a real-valued, continuous, convex function over $\Dist S$. And such vulnerabilities can be expressed by gain functions \cite{Alvim:20a} . A {\bf gain function} is a real-valued function ${\cal W}{\times}S\Fun \Real$, where ${\cal W}$ names (or indexes) the set of actions available to an adversary. Given a gain function $g$ we can define a vulnerability via $V_g(\pi)= \max_{w:{\cal W}} \EV{g(w,-)}{\pi}$. Examples of gain functions have been given in \Sec{s1722}, although there they were as expressions over the program state. So a gain \emph{expression} of the form $\IB{\phi} \,\MAX\, \IB{\psi}$ corresponds to a a gain \emph{function} in which there are two actions ${\cal W}= \{``\phi" , ``\psi" \}$, and
$g(``w", s)$ is $1$ if the state $s$ satisfies $w$, and 0 otherwise.

More interestingly is that --provided the set ${\cal W}$ contains more than one action-- we can measure the {\bf conditional expected vulnerability} over a hyper-distribution as the expected value of $V_g$ (a function on distributions) over $\Delta$ (which is a hyper, i.e.\ distribution of distributions), written $\EV{V_g}{\Delta}$. From standard properties of the expectation operator, we have that $\EV{(-)}{\Delta}$ is continuous, monotone and linear. When $\Delta{=}H(\pi)$ for some {\HMM} $H$ we have that $\EV{V_g}{\Delta}$ is the conditional expected vulnerability over the final state, which means that it takes into account the information leaks as well as the state updates described by $H$.
 
For the program at \Fig{f1544} we can now compute the expected gain wrt.\ the gain expression $\IB{c} \,\MAX\, \IB{\neg c}$ and the final hyper-distribution given at \ref{spade}. For each posterior we compute the maximum expected value with respect to the corresponding gain function and then take the weighted sum of the maxima. This gives $\alpha \times (1\max 0) + (1{-}\alpha)\times (\beta \max (1{-}\beta))$ which is equal to $\alpha + (1{-}\alpha)\times (\beta \max (1{-}\beta))$, as explained at \Fig{f1544}.

\Subsection{Equivalent attacks on the prior}
For any program $H$, prior $\pi$ and vulnerability $V$, it turns out \cite{McIver:15} that there is an equivalent ``pre''-vulnerability  denoted \GPG{H}{V}.  This \emph{pre}-vulnerability  applied to the \emph{input}, the prior, yields the same success as the conditional vulnerability $V$ applied to the \emph{output} hyper-distribution of $H$, that is the following equality holds: $\GPG{H}{V}(\pi)= \EV{V}{H(\pi)}$.  Syntactic rules for determining the prior vulnerabilities, \textbf{our main contribution}, are what we set out set out for \Kuifjetje\ programs in Secs.\ \ref{s1123},\ref{s1124},\ref{s1626} and \Sec{s1626a}.

Properties of $\Gpg$ include well-definedness, that $\GPg{H}$ is linear (and total), and that $\GPg{(H_1;H_2)}= \GPg{H_1}\Comp\GPg{H_2}$, and, of course, that whenever $H_1\HRef H_2$ we must have that $\GPG{H_1}{V_g} \geq \GPG{H_2}{V_g}$. 

%\item
%\underline{
\Subsection{Loops: programs defined by limits}
Programs that contain loops can be modelled as a  limit of unfoldings. Formally these are modelled as ``information flow refinement'' limits used in \QIF\ \cite{Alvim:2012aa,Alvim:20a,McIver:2014aa} and extended for programs in \cite{McIver:11}.
The limits are complicated somewhat if we allow the possibility for non-termination, or even for ``almost sure'' termination, for which the rule below is justified by least fixed points. (Details can be found at \cite{McIver:11}.) In this paper though, all of our loops terminate within a definite finite number of iterations, and in that case a loop can be thought of as a sequential composition of that many unfoldings. For a loop \Exp[LP] defined
\While\ \Exp[G]\ \Do\ \Exp[H]\ \Od\
we have the simple rule that if $[G]{\times}\GPg{H.V_g}= [G]{\times}V_g$ (i.e.\ $V_g$ is an invariant), then we must  also have  $[G]{\times}\GPG{\Exp[LP]}{V_g}= V_g$\,.
 Thus if the advantage (as determined by $V_g$)  to the adversary \emph{after} executing $H$ once is the same as it was \emph{before}  executing it, then the advantage to the adversary of executing the whole loop is the same both before and after as well. \Sec{s1309} separates that into three steps.
 
 %\ToDo{\ldots and therefore is of no use to her.}
 %\footnote{\ToDo {Make this match the actual rule that we use.}}
%\end{enumerate}

\Section{Related work}

The formal analysis of information flow was described first in terms of ``non-interference'' \cite{Goguen:84,Smith:2011aa,Smith07aa}. 
%Smith, Geoffrey (2007). "Principles of Secure Information Flow Analysis". Advances in Information Security. 27. Springer US. pp. 291–307.
The setting makes similar assumptions to those used for \Kuifjetje\ --- an adversary is able to make observations of an executing system and, assuming she has knowledge of the program code, tries to make deductions about some secret. One difference (although not in nature but only of syntax) is that there is a distinct separation between observable program variables and non-observable program variables; the former are called ``high-security variables'' and the latter ``low-security variables'', and a program is said to satisfy \emph{non-interference} if observations of the low-security variables are the same irrespective of the values of the high-security variables. A number of works use type systems to enforce information flow properties, including \cite{Sabelfeld:03,DBLP:conf/csmr/LiuM10,ONeill:06}. Information flows wrt.\ other threat models have been surveyed  comprehensively by Kozri et al.\ \cite{Kozyri:22}.

Non-interference has been found to be excessively severe or cannot be implemented, and there are many examples of  practical systems that do not satisfy it.  Qualitative models which do permit \emph{some} information flows include Sabelfeld and Sands \cite{Sabelfeld:01},  Morgan \cite{Morgan:06}, and some quantitative models were also introduced by Aldini \cite{Aldini:04}, though do not include the $g$-leakage framework that we use here.

An important generalisation of these  is Quantitative Information flow which allows for some information leaks so that the results of a  verification exercise shifts from asserting that there are no information leaks to admitting some but then measuring their extent and impact. \QIF\ models have been proposed in various forms for example for deterministic programs Clarke et. al \cite{DBLP:journals/jcs/ClarkHM07}, Heusser and Malacaria \cite{Heusser:10} for  statistical databases with passive adversaries and K\"opf and Basin for side channels with adaptive adversaries \cite{Kopf:07}.  
% Boris Köpf and David Basin. 2007. An Information-theoretic Model for Adaptive Side-channel Attacks. In Proceedings of the 14th ACM Conference on Computer and Communications Security (CCS '07). ACM, New York, NY, USA, 286–296.
 Smith's pioneering work \cite{Smith:2009aa}  led to the realisation that measurements based on traditional information theory such as Shannon entropy did not give an accurate assessment of the impact of information leaks --- including the intent of the adversary is crucial. This approach has been explored in many works such as Alvim et al.\ \cite{Alvim:2012aa,Alvim:20a} for security generally, and more specifically for privacy \cite{DBLP:phd/hal/Fernandes21,DBLP:journals/jcs/AlvimACDP15}. One important outcome of the approach is the discovery of a way to compare systems using the partial order of refinement used here.  

The use of refinement for reasoning about security in traditional programming settings led to the ``refinement paradox'' \cite{Jacob:88} where apparently secure programs had refinements that were insecure in the sense that an adversary could discern a secret value in and implementation that was prevented in the specification). Morgan's Shadow semantics \cite{Morgan:06} based on an adaptation of Kripke models, demonstrated how to avoid the paradox. One of its innovations was to expose the necessity for implicit flows to ensure important principles of compositionalty. Other explorations of refinement-style reasoning for security include  Hunt and Sands  \cite{Hunt021}.

The first example of program semantics and backwards reasoning for probabilistic programs was proposed by Kozen for deterministic programs (based on an underlying Markov model for updates) \cite{Kozen:81}; it was generalised by McIver and Morgan to include demonic nondeterminism (using \MDP's, Markov \emph{decision process} for updates) \cite{McIver:05a}, then further explored by Gretz et al.\ \cite{GretzKM14} and Kaminski \cite{Kaminski19}.  The introduction of Bayesian updates to incorporate information flow (based on a Hidden Markov Model) was introduced by Meinicke et al.\ \cite{mcivermeinicke10a,McIver:15}, but was also found in Jansen  et al.\ \cite{0001KKOGM15}.

A compiler for the \Kuifjetje\ language was produced originally by Schrijvers et al.\ \cite{FoPPS:19}; implemented in Haskell, it makes heavy use of the Probabilty Monad \cite{Giry:81,Jones:89}. It computes the hyper-distribution output of a program enabling gain-function style experiments to be performed, and was used by Jurado and Smith for their analysis of searchable encryption \cite{JuradoP021}. Although the \Kuifjetje\ compiler can produce finite representations for certain scenarios, it does not have the capability of proving information-flow properties generally, which is our primary contribution here. Interestingly, recent work by Kornaropoulos et al.\ \cite{Kornaropoulos:22} has looked at leakage properties of searchable encryption with their analysis being similar to ours in terms of trying to identify properties of databases that would leave them vulnerable wrt.\ relevant leakage measures.

Hidden Markov Models were introduced by Baum et al.\ \cite{Baum:70}; and model-checking  for formal analysis of them have been explored by Zhang et al.\ \cite{Zhang2005LogicAM} and Buckby et al.\ \cite{Buckby:20}. 

\Section{Conclusion}

We have presented what we believe to be the first \emph{source-level} reasoning framework for reasoning about \QIF\ in programs having a (relatively new) hyper-distribution semantics taken from \cite[\S2.2]{mcivermeinicke10a}\cite[\S4.3]{Alvim:20a}; the mini- programming language \Kuifjetje, corresponding to it, is lifted from \cite[Part IV]{Alvim:20a}. 
Other extensive studies of leakage analysis are typically in terms of pseudo-code related informally to the precise mathematical semantics whose properties are of interest.\,%
%Although others have studied leakage analysis extensively, for decades, it is usually in terms of pseudo-code related informally to the precise mathematical semantics studied.\,%
\footnote{For programs this appears to be the first to apply \emph{quantitative} formal reasoning directly at the source level for information flow. Joshi and Leino \cite{Leino:00} have however studied source-level reasoning for \emph{non-quantitative} information flow, as has Morgan \cite{Morgan:07}.}
%\ToDo{Are we sure of that, even for non-hyper semantics?}.}\

{\bf Our principal contribution} is a \underline{formal} interpretation of \emph{leakage specifications}, Hoare-triple-style annotations
\[
	\{\,\Exp[preGain]\,\} ~~ \Exp[prog] ~~ \{\,\Exp[postGain]\,\}
\]
that abstract, as formal programming logics do, from the mathematical semantics of the programs themselves (\HMM's in our case). Validity of the above in our logic means that the expected gain of the expression \Exp[postGain], accruing to the adversary over the \underline{out}p\underline{ut} state of \Exp[prog] (a real number, which we have written with \$'s), is equal to the expected gain of \Exp[preGain] over the prior distribution on the \underline{in}p\underline{ut} state from where the program was run (according to its \HMM\ semantics).

The significance of \Exp[postGain]'s expressiveness is that very sophisticated descriptions of adversaries' potential attacks, and their expected gains, can pose quite precise questions about a system's security in a \emph{variety} of environments --- not just the ``Can she guess?''\ scenarios we have presented here (\App{a1049}).

The significance of ``source level'' is that it promotes more reliable reasoning, even on paper --- recall the plausible but \emph{incorrect} analysis \Eqn{e1949} of \Fig{f1544}. But much more importantly is its potential for mechanisation in systems such as Coq \cite{Coq} or Isabelle \cite{Isabelle}. That would enable, in the future, leakage analysis to be applied to programs in the same way that proofs of cryptographic software can be achieved currently \cite{Barthe:12}.

Especially important for this kind of analysis is that the proofs can be achieved in a compositional manner, suggesting the possibility for re-use and extension of proofs in larger program contexts.

\newpage
%\bibliographystyle{plain}
%\bibliography{probsNew230130}

\appendices
\cleardoublepage
\Section{Expressiveness of gain functions/expressions\Label{a1049}}

The post-gain in \Fig{f1356} expresses that the adversary will gain \$1 if from running the program she learns where $x$ is in $A$ --- but if $x$ is not in $A$, she gains nothing, i.e. \$0. The given post-gain, by omitting\quad\MAX\,\IB{\,x{\notin}A\,}\quad does not allow her to profit from learning that $x$ is not in $A$ at all.

Those gains, expressed as \$-values, are only abstractions from \emph{actions} the adversary might take with that knowledge.

So now suppose that $A[0{:}N]$ represents a row $A[0]$\,--to--\,$A[N\!{-}1]$ of warehouses each of which might have a \$1 coin inside: let $A$ have type Boolean; and let $x$ be \True, meaning ``Yes, there's \$1 in here.'' Then our post-gain is saying that the adversary will gain \$1 if she raids a warehouse with a coin inside.\,%
\footnote{This can easily be made more realistic by using \$1,000,000 instead of \$1. And what might we do if the warehouses have treasures of varying value? There are many further  possibilities.}

If $A$ contains no coins: as the post-gain is written, she gains nothing in that situation (as we mentioned above). But perhaps she could sell that knowledge to another adversary, a partner in crime, saying ``Don't raid $A$ --- there's nothing there. Try raiding $B$ instead.'' If giving that advice is worth say 10\Cents\ to \emph{our} adversary, because the \emph{other} adversary will pay her for it, then we would use a different post-gain in \Fig{f1356}, that is
\[
	\GFQ{i}{0{\leq}i{<}N}{\IB{\,A[i]{=}x\,}} \MAX\ \NF{1}{10}\IB{\,x{\not\in}A\,}\,
\]
to describe our (now more enterprising) adversary. The resulting pre-gain would then become \IB{\,x{\in}A\,}\ \MAX\ \NF{1}{10}\IB{\,x{\not\in}A\,}\,, meaning ``If I find a coin, I'll steal it myself; but if I learn there's no coin anywhere, I'll sell that knowledge for 10\Cents\,.''

Finally, suppose our adversary's policy is ``Raid a particular $A[n]$ if she knows there's a coin there; but if she knows only that there's a coin in $A$ \emph{somewhere}, then raid them all at the same time and take one coin --- but absorb a cost of 40\Cents\ for the extra equipment she needed for the full assault. And if she knows there's no coin, then sell that information (as before).
We would describe her with the post-gain
\[
	\GFQ{i}{0{\leq}i{<}N}{\IB{\,A[i]{=}x\,}}\; \MAX\ \NF{3}{5}\IB{\,x{\in}A\,}\; \MAX\ \NF{1}{10}\IB{\,x{\not\in}A\,}\,,
\]
where \NF{3}{5} is $\$1{-}40\Cents$.
Figures \ref{f1541a},\ref{f1541b} show further possibilities.

\begin{figure}
\begin{tabular}{lr}
$\{{\MAx}\,A\}$ &\Cmt{Pre-gain} \\[0.5ex]
$n,g$:= $0,0$ &\Cmt{$g$ is greatest in $A[{:}n]$} \\
\While\ $n{\neq}N$ \Do \\
\quad	\If\ $A[n]{>}g$ \Then\ $g$:= $A[n]$ \\
\quad $n$:= $n+1$ \\
\Od \\
\multicolumn{2}{l}{\Cmt{Note there's no ``{\small\Print}\ $g$''.}}\\
\multicolumn{2}{l}{\Cmt{The (implicit) leak comes from the test $A[n]{>}g$.}}\\[0.5ex]
\GFQ{i}{0{\leq}i{<}N}{A[i]} &\Cmt{Post-gain}
\end{tabular}

\bigskip
Note that because $g$ is not \Print'ed, the adversary learns \emph{where} the most ``money'' is; but not \emph{how much} it is. Even so, she gets that maximum.

\medskip Note also that the post-gain does not have \IB{-}'s, because we are now dealing with properly numeric values, no longer only ``whether or not'' Booleans converted to nominal \$1 or \$0.
\caption{Where is the most money?\Label{f1541a}}
\end{figure}

\begin{figure}
\begin{tabular}{lr}
\GFQ{i}{0{\leq}i{<}N}{A[i]} &\Cmt{Pre-gain} \\[0.5ex]
$n,g$:= $0,0$ \\
\While\ $n{\neq}N$ \Do &\Cmt{$g$ is greatest in $A[{:}n]$} \\
\quad	$g$:= $g\sqcup A[n]$ \\
\quad $n$:= $n+1$ \\
\Od \\
\GFQ{i}{0{\leq}i{<}N}{A[i]} &\Cmt{Post-gain}
\end{tabular}

\bigskip
Here --by removing the leaky \If\ from \Fig{f1541a}-- we ensure that the adversary learns nothing at all about $A$,\,\footnotemark\ 
and so her pre-gain is the same as the post-gain, i.e.\ is based entirely on her prior knowledge of $A$'s distribution. Thus her strategy can only be to choose $n$ such that the expected value of $A[n]$, over $A$'s p\underline{rior}, is greatest.
\caption{Second example for \App{a1049}\Label{f1541b}}
\end{figure}
\footnotetext{Determining $N$ by counting iterations as the loop runs tells her nothing she doesn't already know from the prior.}

\end{document}